\documentclass[10pt, preprint2]{aastex}

\begin{document}

\title{Interesting Evidence for a Low-Level Oscillation Superimposed on the Local Hubble Flow}

\author{M.B. Bell\altaffilmark{1}}
%and S.P. Comeau\altaffilmark{1}}

\altaffiltext{1}{National Research Council of Canada, 100 Sussex Drive, Ottawa,
ON, Canada K1A 0R6; morley.bell@nrc.gc.ca}
%\altaffiltext{2}{Dept of Physics, University of Guelph, Guelph, ON, Canada, N1G %2W1; comeaus@uoguelph.ca}

 \begin{abstract}

Historically the velocity scatter seen on local Hubble plots has been attributed to the peculiar velocities of individual galaxies. Although most galaxies also have uncertainties in their distances, when galaxies with accurate distances are used recent studies have found that these supposed peculiar velocities may have preferred, or discrete, values. Here we report the interesting result that when these discrete components are identified and removed from the radial velocities of the SNeIa galaxies studied in the Hubble Key Project, there is evidence for a residual oscillation, or ripple, superimposed on the Hubble flow. This oscillation has a wavelength near 40 Mpc and, because its amplitude is small compared to that of the scatter in velocities, it becomes visible only after the discrete components are removed. This result is interesting because even if this ripple has been produced by a selection effect, the fact that it is only revealed after the discrete velocities are removed implies that the discrete velocities are real. Alternatively, if no selection effect can be identified to explain the ripple, then both the discrete velocities and the ripple together become very difficult to explain by chance and these results could then have interesting cosmological consequences.

\end{abstract}

%\keywords{galaxies: active - galaxies: distances and redshifts - galaxies: %quasars: general}

\keywords{galaxies: Cosmology: distance scale -- galaxies: Distances and redshifts - galaxies: quasars: general}

%\maketitle

\section{Introduction}

It has been demonstrated \citep{tif96,tif97}, (and related papers) that there appear to be discrete "velocity periods" present in the redshifts of galaxies. The most obvious of these was found in common spirals and showed discrete velocity components near 36, 72, 145, 290, etc., km s$^{-1}$. In each of the other period groups detected, the velocities showed this same octave related, or doubling nature. More recently we have found evidence that the extrapolation of Tifft's periods to higher components, using this doubling relation, leads to discrete velocity components that appear to be visible in the radial velocities of all galaxies whose distances are accurately known \citep{bel03a,bel03b,rus03,rus04}.
%Evidence has also been reported suggesting that there may be even larger %discrete intrinsic redshift components present in QSO redshifts %\citep{bel02a,bel02b,bel02c,bel02d,bel04,bel06,mcd06,mcd07,bel07}. 

These small non-cosmological redshift components in galaxies introduce a scatter in the Hubble plot that is much larger than can be explained by the errors of measurement. This scatter has been explained historically by peculiar velocities, although in some cases this has been questioned simply because of their large size \citep{dav05a,dav05b,dav05c}. If the quantization found in these velocities is real it would appear to rule out the peculiar velocity interpretation.

In this paper we report a remarkable result in which a low-level ripple is clearly seen to be present in the residual Hubble plot after the discrete velocities are identified and removed from the velocities of the SNeIa galaxies studied in the Hubble Key Project. Because its amplitude is small compared to the scatter in velocities it is not visible before the discrete velocity components are removed. One model that might explain this type of oscillation in the Hubble flow has been discussed previously by \citet{mor90,mor91}.

\section{Discrete Velocities Defined in Previous Work}

The discrete "velocity" components in galaxies, found by \citet{tif96,tif97} and recently confirmed by us, can be expressed by the relation:

%\begin{equation}

z$_{iG}$[$N,m$] = (z$_{iG}$[$N,n_{max}$]/2$^{m}$)$_{m=0,1,2,3..\infty}$ ----(1)

%z_{iG}[N,m] = %\left(\frac{z_{iQ}\left[N,n_{max}\right]}{2^{m}}\right)_{m=0,1,2,3..\infty}

%\end{equation}

Here the integral values of the quantum number $N$ correspond to the different "velocity" periods identified by Tifft, and the quantum number $m$ represents the number of halvings (via 2$^{-m}$) below the relevant maximum intrinsic redshift component z$_{iQ}$[$N,n_{max}$] in each $N$-group \citep{bel03a,bel03b}. The maximum for galaxies is equal to the minimum found previously for quasars in each relavent group \citep{bel02a,bel02b,bel02c,bel02d,bel07}.

Until a physical explanation of the discrete components becomes known and it is determined if these are discrete velocities as suggested by Tifft, or preferred redshifts, in which case they might be related to the atom itself, or even some systematic effect, it is not possible to explain what the $N$ and $m$ quantum numbers might be due to.

The discrete components in galaxies get quite small as $m$ increases and because of this cannot be resolved at high $m$-values. Values for the larger (low-$m$) discrete redshift components in galaxies, and their velocity equivalents, obtained using several independent galaxy groups containing 138 galaxies with accurate Tulley-Fisher distances, are listed in \citet[Table 4]{bel03b} for $N$ = 1 to 6. For each value of $N$ the relevant z$_{iQ}$[$N,n_{max}$] value is given by the highlighted value in \citet[Table 2]{bel02d}. The period group found by Tifft to be associated with common spiral galaxies, corresponds to the lowest, $N$ = 1, group of discrete velocities.

Because we looked at galaxies that were more distant than those studied by Tifft we were also able to conclude that the discrete components are superimposed on top of the Hubble flow.

\section{Analysis of Type IA Supernovae Data}

A preliminary analysis of the SNeIa data was reported previously \citep{bel03b} and it followed the analysis used for spiral galaxies \citep{bel03a,bel03b} where a minimum was sought in the RMS deviation in source velocities, calculated relative to the nearest discrete velocity line superimposed on the Hubble flow. This analysis technique is explained in more detail by \citet{rus03,rus04}.

In Fig 1 the RMS deviations in V$_{\rm CMB}$ velocities, relative to the nearest discrete velocity line in Fig 2, are plotted vs H$_{\rm o}$, using the 36 SNeIa galaxies listed in Table 6 of \citet{fre01}. The shape of this curve and a demonstration on how the dip at H$_{\rm o}$ is produced when discrete components are present are discussed in detail by \citet{rus04}. A clear best-fit feature is visible here at H$_{\rm o}$ = 58 km s$^{-1}$ Mpc$^{-1}$, as was found for the 138 spiral galaxies studied previously \citep{bel03b,rus03}. We have shown previously \citep{bel03a,rus03} that there is nothing in our analysis procedure that can always produce an RMS dip at the same value of H$_{\rm o}$ = 58 in random data. The fact that a similar dip at H$_{\rm o}$ = 58 was found in the several independent galaxy groups we looked at therefore makes this result very difficult to explain by chance. The minimum at H$_{\rm o}$ = 67 in Fig 1 is due to the overall shape of the V$_{\rm CMB}$ vs Distance source distribution. It is not produced by the presence of intrinsic components and appears in all source distributions, even randomly generated ones where there are no discrete velocity components present. It can be predicted to be slightly lower than the value that would be obtained by fitting a single straight line to the data (as is done in the Hubble Key Project which found a value of H$_{\rm o}$ = 72).

 In Fig 2, the V$_{\rm CMB}$ velocities of the 36 SNeIa galaxies are plotted vs distance. The slope of the discrete velocity lines is determined by H$_{\rm o}$ and their discrete velocity values are defined by equation 1 above. Here it can be seen that the RMS dip at H$_{\rm o}$ = 58 is obtained when several sources in each $[N,m]$-group fall along the discrete redshift lines over an extended range of distances. This is particularly obvious in Fig 2 for the [$N,m$] = [2,5], [5,8] and [5,9] lines, where 23 of the 32 sources below 300 Mpc fall along these three lines. Because of this it is possible to identify which $N$-group the sources are from. Unlike the other groups of galaxies studied, at least 60 percent of these sources appear to be $N$ = 2 sources, while the remaining 40 percent are $N$ = 5 sources. At the same time these sources lie at much greater distances, starting near 60 Mpc, where the previously studied groups leave off, and extending to 450 Mpc. The N values may relate to the type of galaxy that generates Type Ia SNe in the redshift range studied here. Tifft studied closer galaxies and found that N = 1 and 2 were most common. However, until an explanation for the discrete components is found this can only be speculation.

It is also interesting to note that in some respects the plot in Fig 2 is similar to that in Fig 2 of \citet{bel07}, where intrinsic velocity or redshift components appear to increase with more distant objects.

\begin{figure}[h,t]
\hspace{-1.0cm}
\vspace{-1.8cm}
\epsscale{0.9}
\includegraphics[width=9cm]{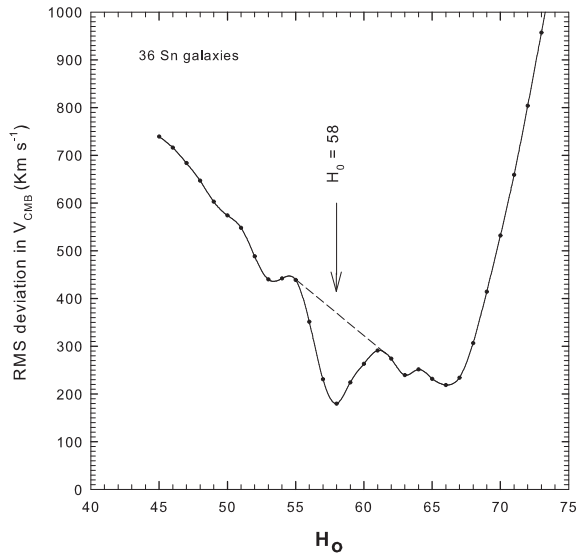}
%\includegraphics[width=9cm]{rmsplotn25a.eps}
%\plotone{fig1.eps}
\caption{{RMS deviation in V$_{\rm CMB}$ about the intrinsic redshift grid lines in Fig 2 vs H$_{\rm o}$ for Type IA Supernovae. Data are from \citet{fre01} \label{fig1}}}
\end{figure}
 
\begin{figure}[h,t]
\hspace{-0.8cm}
\vspace{-1.0cm}
\epsscale{0.9}
\includegraphics[width=9cm]{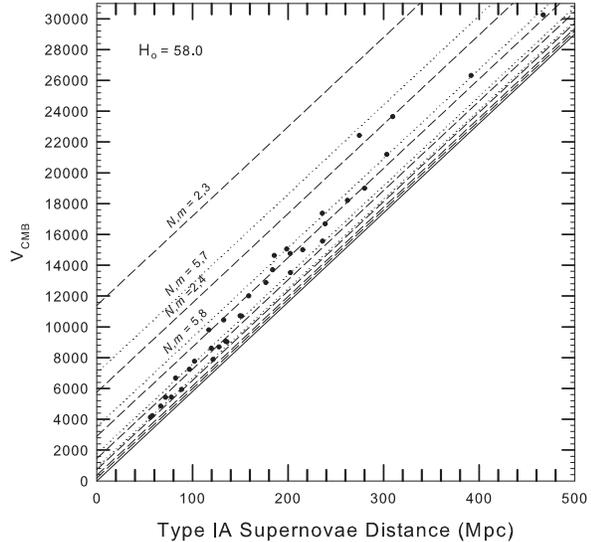}
%\plotone{fig2.eps}
%\plotone{sn1mb5.eps}
\caption{{Hubble plot for Type IA Supernovae galaxies. Data are from \citet{fre01}. Dashed lines are $N$ = 2 group and dotted lines are $N$ = 5 group. \label{fig2}}}
\end{figure}

\begin{figure}[h,t]
\hspace{-1.0cm}
\vspace{-1.5cm}
\epsscale{0.9}
\includegraphics[width=9cm]{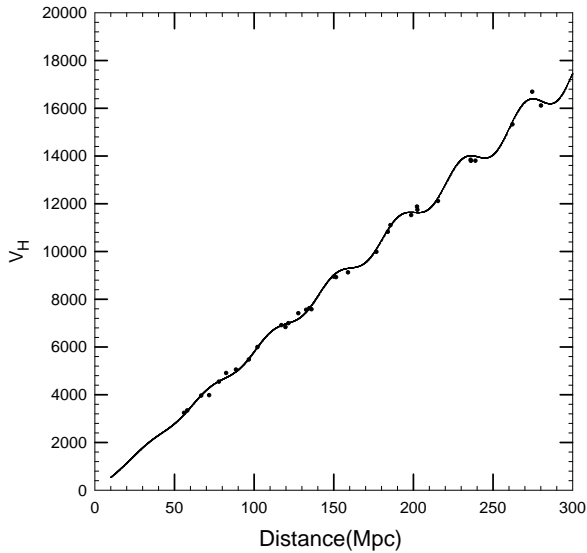}
%\plotone{fig3.eps}
%\plotone{rip1b.eps}
\caption{{Plot of Hubble velocity as a function of Distance for SNeIa galaxies after removal of intrinsic redshifts. \label{fig3}}}
\end{figure}
 
\begin{figure}[h,t]
\hspace{-0.8cm}
\vspace{-1.5cm}
\epsscale{0.9}
\includegraphics[width=9cm]{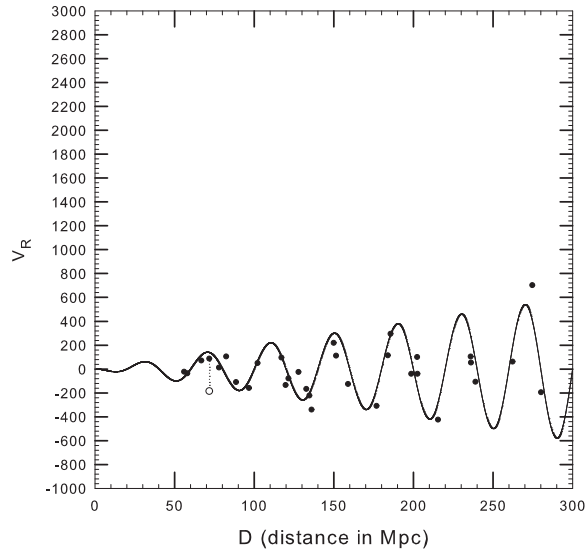}
%\plotone{fig4.eps}
%\plotone{vhslope40a.eps}
\caption{{Plot of residual velocity as a function of distance for SNeIa galaxies after removal of intrinsic redshifts and Hubble slope of 57.9 km s$^{-1}$ Mpc$^{-1}$. Redshift peaks are located at 30, 70, 110, 150, 190, 230 and 270 Mpc. \label{fig4}}}
\end{figure}
 
\begin{figure}[h,t]
\hspace{-0.8cm}
\vspace{-1.0cm}
\epsscale{0.9}
\includegraphics[width=8cm]{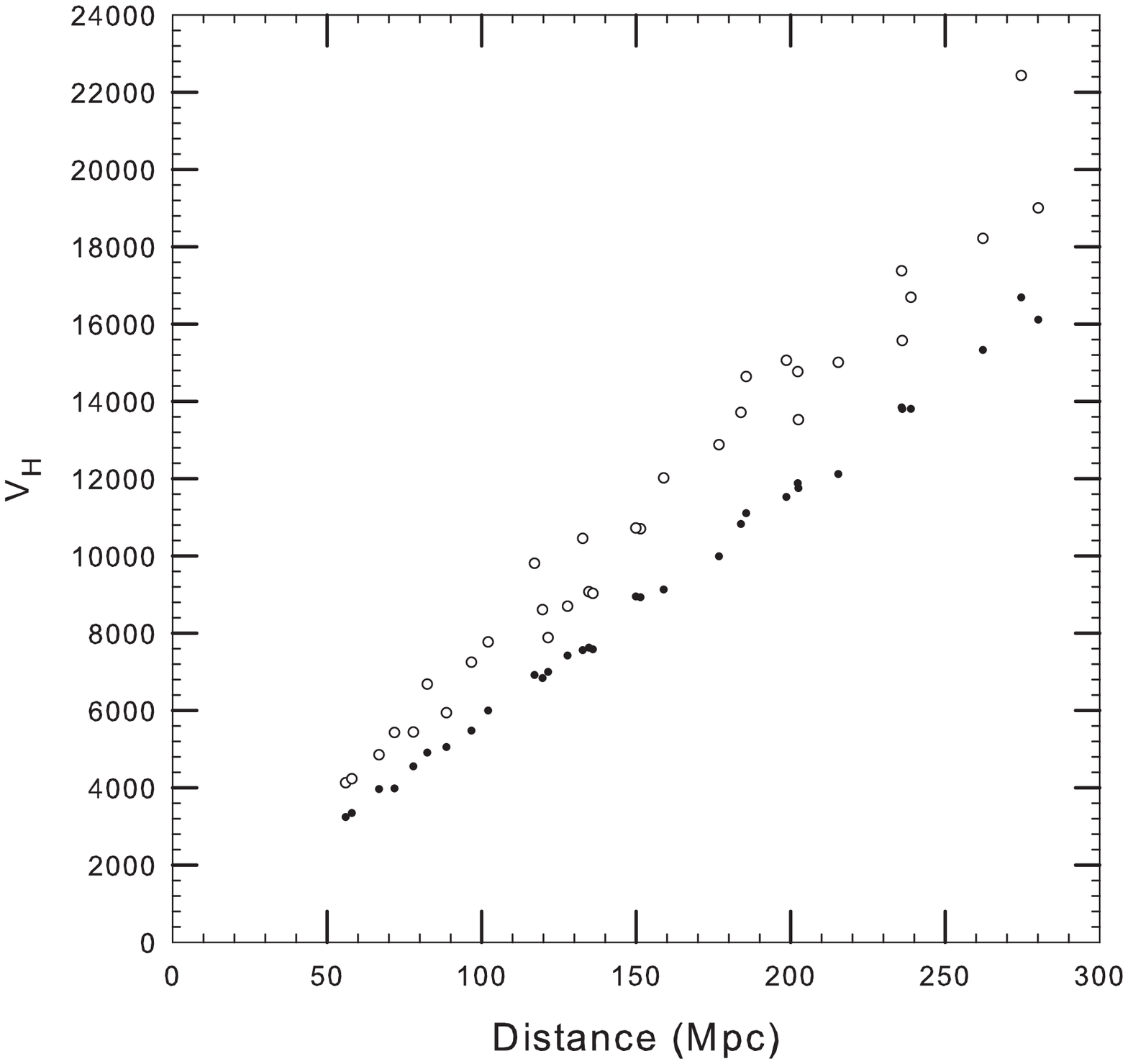}
%\includegraphics[width=8cm]{vcmbvsd4.eps}
%\plotone{fig2.eps}
%\plotone{sn1mb5.eps}
\caption{{V$_{\rm CMB}$ plotted versus Distance before (open circles) and after (filled circles) discrete components are removed. As can be seen, the ripple amplitude is too small to be detected before the intrinsic values are removed. \label{fig5}}}
\end{figure}

\section{Hubble Plot After Removal of Discrete Velocity Components}

Table 1 lists the SNeIa sources in col 1, with their distances and V$_{\rm CMB}$ velocities in cols 2 and 3 respectively. Col 4 gives the intrinsic redshift [$N,m$] (and its associated "discrete velocity") obtained for each source in the best-fit situation for H$_{\rm o}$ = 58. Col 5 lists the Hubble velocity V$_{\rm H}$ after removal of the discrete velocities.

In Fig 3 the Hubble velocities (V$_{\rm H}$) are plotted vs distance. A linear regression on the data (all 36 points) gave a slope of 58.26 (std. err. 0.33) km s$^{-1}$ Mpc$^{-1}$. Because there are only four sources between 300 and 450 Mpc, and because the distance uncertainties may increase in proportion to distance, only those 32 sources closer than 300 Mpc have been considered for further analysis. It is important to understand here that the problem related to the increasing distance uncertainty is not one of fitting to the ripple, whose amplitude may also increase with distance, but one of fitting to the discrete velocity lines whose separations do not increase. When the distance uncertainties approach the spacing between these lines the result becomes ambiguous. 

Although there are many more SNeIa galaxies with redshifts between z = 0.4 and 0.8 measured in the Supernova Cosmology Project \citep{per99}, these have not been included because they lie at redshifts where cosmological corrections need to be taken into account, and these can be confused with the intrinsic redshifts considered here.

%unfortunately there is a large gap between those sources and the low redshift %(z $< 0.1$) sources studied in this paper. Those sources have not been examined %here for several reasons. Since they are located at significantly larger %distances, their distance uncertainties may be too large to permit unambiguous %fitting to the grid lines, and their distances are given as magnitudes only. %However, the main reason 

\begin{table*}
\caption[]{Parameters for Supernovae Galaxies.}
\label{tab-1}
$$
\begin{tabular}[]{ccccc}
%\begin{array}{ccccc}
\hline
Supernova & D (Mpc) & V$_{\rm CMB}$ (km s$^{-1}$)[1] & Transit.(Disc.Vel.)(km s$^{-1}$)[2] & V$_{\rm H}$ (km s$^{-1}$)[3] \\

\hline

SN 1990O  & 134.7  & 9065  & z$_{\rm iG}$[2,6](1448)    & 7617  \\ 
SN 1990T  & 158.9  & 12012 & z$_{\rm iG}$[2,5](2890)    & 9122  \\ 
SN 1990af & 198.6  & 15055 & z$_{\rm iG}$[5,8](3537)    & 11518  \\
SN 1991S  & 238.9  & 16687 & z$_{\rm iG}$[2,5](2890)    & 13797 \\ 
SN 1991U  & 117.1  & 9801  & z$_{\rm iG}$[2,5](2890)    & 6911 \\ 
SN 1991ag & 56.0   & 4124  & z$_{\rm iG}$[5,10](888)    & 3236 \\ 
SN 1992J  & 183.9  & 13707 & z$_{\rm iG}$[2,5](2890)    & 10817  \\ 
SN 1992P  & 121.5  & 7880  & z$_{\rm iG}$[5,10](888)    & 6992 \\ 
SN 1992ae & 274.6  & 22426 & z$_{\rm iG}$[2,4](5742)    & 16684  \\ 
SN 1992ag & 102.1  & 7765  & z$_{\rm iG}$[5,9](1772.5)  & 5992  \\ 
SN 1992al & 58.0   & 4227  & z$_{\rm iG}$[5,10](888)    & 3339 \\ 
SN 1992aq & 467.0  & 30253 & z$_{\rm iG}$[2,5](2890)    & 27363  \\
SN 1992au & 262.2  & 18212 & z$_{\rm iG}$[2,5](2890)    & 15322 \\ 
SN 1992bc & 88.6   & 5935  & z$_{\rm iG}$[5,10](888)    & 5047 \\ 
SN 1992bg & 151.4  & 10696 & z$_{\rm iG}$[5,9](1772.5)  & 8923 \\ 
SN 1992bh & 202.5  & 13518 & z$_{\rm iG}$[5,9](1772.5)  & 11745 \\ 
SN 1992bk & 235.9  & 17371 & z$_{\rm iG}$[5,8](3537)    & 13834 \\ 
SN 1992bl & 176.8  & 12871 & z$_{\rm iG}$[2,5](2890)    & 9981 \\ 
SN 1992bo & 77.9   & 5434  & z$_{\rm iG}$[5,10](888)    & 4546 \\  
SN 1992bp & 309.5  & 23646 & z$_{\rm iG}$[2,4](5742)    & 17904 \\ 

SN 1992br & 391.5  & 26318 & z$_{\rm iG}$[5,8](3537)    & 22781 \\
SN 1992bs & 280.1  & 18997 & z$_{\rm iG}$[2,5](2890)    & 16107 \\
SN 1993B & 303.4   & 21190 & z$_{\rm iG}$[5,8](3537)    & 17653 \\ 
SN 1993O & 236.1   & 15567 & z$_{\rm iG}$[5,9](1772.5)  & 13795 \\
SN 1993ag & 215.4  & 15002 & z$_{\rm iG}$[2,5](2890)    & 12112\\ 
SN 1993ah & 119.7  & 8604  & z$_{\rm iG}$[5,9](1772.5)  & 6832 \\ 
SN 1993ac & 202.3  & 14764 & z$_{\rm iG}$[2,5](2890)    & 11874 \\ 
SN 1993ae & 71.8  & 5424   & z$_{\rm iG}$[2,6](1449)[1,4](1158) & 3975(4266) \\ 
SN 1994M & 96.7   & 7241   & z$_{\rm iG}$[5,9](1772.5)  & 5469 \\  
SN 1994Q & 127.8  & 8691   & z$_{\rm iG}$[2,6](1449)    & 7413 \\ 
SN 1994S  & 66.8  & 4847   & z$_{\rm iG}$[5,10](888)    & 3959 \\  
SN 1994T & 149.9  & 10715  & z$_{\rm iG}$[5,9](1772.5)  & 8943 \\ 
SN 1995ac & 185.6 & 14634  & z$_{\rm iG}$[5,8](3537)    & 11097 \\ 
SN 1995ak & 82.4  & 6673   & z$_{\rm iG}$[5,9](1772.5)  & 4901 \\
SN 1996C & 136.0  & 9024   & z$_{\rm iG}$[2,6](1449)    & 7575 \\ 
SN 1996bl & 132.7  & 10446 & z$_{\rm iG}$[2,5](2890)    & 7556 \\

\hline
\end{tabular}
%\end{array}
$$
%\end{table*}
[1] Total measured velocity including discrete component.
 
[2] Discrete component obtained.
  
[3] Hubble velocity after removal of discrete component.

\end{table*}

In Fig 3, where the velocities no longer contain discrete intrinsic components, there is now evidence for a low-level modulation superimposed on top of the otherwise linear Hubble slope. It has been approximated by the sinusoid (solid curve) to which all sources can be reasonably well fitted.

Since there is nothing in our analysis that could have produced this modulation, it appears to indicate that there is a low-level, sinusoidal oscillation in the expansion rate of the local Universe. Changes in the expansion rate are not new, since it is now well accepted, at least over a much longer time scale, that the expansion rate of the Universe is currently accelerating. Investigators have argued that ripples in the Hubble flow might someday be detected \citep[see for example]{dav04} and others have already reported periodic density clumping with redshift \citep{bro90,moh92,har08} (see below for further discussion).

In Fig 4, a slope of 57.9 has been removed from the V$_{\rm H}$ vs distance plot in Fig 3. Two velocities have been plotted for SN 1993ae, which is located at a distance of 71.8 Mpc. The open circle is the velocity obtained if this source is fitted to the nearest $N$ = 2 group ($N,m$ = 2,6). For this assignment it clearly does not fit the curve. However, if this source is fitted to the nearest grid line ($N,m$ = 1,4), it fits the curve well. It is therefore assumed that this one source is an $N$ = 1 group object.

Fig 4 clearly shows the presence of a sinusoidal fluctuation in the residual velocity (V$_{\rm R}$) with increasing cosmological distance. The best-fit wavelength of this modulation is $\lambda_{g}$ = 39.6 Mpc (period = $\sim1.3\times10^{8}$ yrs, frequency $\nu_{g} \sim2.4\times10^{-16}$ Hz). Its best-fit phase for a sine wave with 39.6 Mpc wavelength is 3.18 radians, with redshift peaks at 29.7, 69.7, 109.7, etc., $\pm0.3$ Mpc. We also note that the sources used here are located in all directions on the sky, confirming the isotropic nature of these results. The amplitude of the oscillation may also increase with distance, at least initially.

The observed ripple, with a wavelength near 40 Mpc, is similar to one of the periods reported previously by \citet{har08}. Their distance spacing of (31.7$\pm1.8)$h$^{-1}$ corresponds to 44$\pm8$ Mpc for H$_{\rm o}$ = 72$\pm8$ (h = 0.72$\pm0.08$), which is the relevant value for H$_{\rm o}$ to be used if the discrete components have not been removed. The spacing they find between peaks for this component is thus in good agreement with our 40 Mpc ripple.

The ripple cannot be detected before the intrinsic components have been removed, but it can easily be seen by eye after the discrete components are removed, both before and after the Hubble slope is removed (Figs 3 and 4). There is no obvious sign of this ripple before the discrete components are removed because the scatter in velocities before removal of the discrete components is significantly larger than it is after removal of the discrete components. This is demonstrated more clearly in Fig 5 where V$_{\rm CMB}$ is plotted versus distance, both before and after removal of the intrinsic components. There is no obvious sign of the ripple before the intrinsic components are removed.
 
It is important to point out that to our knowledge no one has previously reported a velocity oscillation. Previous results using much larger samples have found density ripples from a periodic clumping with redshift \citep{bro90,moh92,har08}. When a velocity oscillation is found there is every reason to suspect that there might also be density clumping with the same period. This was pointed out by Morikawa (1990), who states that an oscillation of the Hubble parameter will cause an apparent density fluctuation. If there is a density ripple present with a similar period in other source samples its detection would not have been prevented by the presence of discrete velocity components if the distances are accurately known.  

The presence of intrinsic redshift components in the radial velocities of galaxies (velocity components that lie only on the high velocity side of the Hubble flow as here) will result in a Hubble constant that is too high if they are not taken into account. This has led to a suggested, revised value for the Hubble constant of H$_{\rm o}$ = 58 km s$^{-1}$ Mpc$^{-1}$ \citep{bel03b,rus03}. This is $\sim20$\% lower than the value reported by the Hubble Key Project, but in good agreement with the value found for intermediate epoch galaxies using the Sunyaev-Zel'dovich effect \citep{jon01,mas01,ree02,ree03}.

\section{Do Selection Effects Play a Role?}

If some identifiable selection effect can be found that would produce an oscillation in velocity that increases in amplitude as a function of distance it would be interesting because it would mean that since the removal of the discrete components is necessary before the ripple becomes visible, the discrete components are then unlikely to be random. If no selection effect can be found it also makes the ripple very difficult to explain if it is not real since it would then require that two random components have worked together in just the right way to produce these results by chance, which would be highly unlikely.

\section{Discussion}

There are several questions that need to be addressed before concluding our investigation. Does the presence of the ripple in the raw V$_{\rm CMB}$ data produce the dip seen in Fig 1? Does its presence affect the dip? Is the ripple somehow produced by the reduction procedure?

The data analysis used here contains three main steps as follows:

1) The best RMS fit of the data to the nearest intrinsic line is calculated as the Hubble slope in Fig 2 is varied in unit steps from 45 to 75. These values are plotted versus H$_{\rm o}$ where an RMS dip will be detected when data points align with intrinsic components (Fig 1). 

2) The discrete components obtained for the best-fit H$_{\rm o}$ = 58 results are subtracted from the V$_{\rm CMB}$ values to obtain the result in Fig 3. This involves a simple linear subtraction that should not produce a ripple.

3) The Hubble slope is subtracted from the data in Fig 3 to obtain the residual velocity V$_{\rm R}$ plotted in Fig 4. Again this involves a simple linear subtraction that should not produce a ripple.

To investigate whether or not the data reduction process produced the ripple we repeated the reduction process after first removing the ripple in Fig 4 from the raw V$_{\rm CMB}$ values. This time no ripple was detected. The ripple was then subtracted from the data in Fig 4 and the result was compared to the result obtained from the second processing. The two were found to be identical, as expected. This showed that the ripple is only found when one is present in the data and we can conclude that the ripple is a real feature that was not created by our analysis procedure.

However, it is also important to show that it was not the presence of the ripple in the raw data that produced the dip seen at H$_{\rm o}$ = 58 in Fig 1. If the ripple is removed from the raw V$_{\rm CMB}$ data before processing is the dip affected? Does it disappear completely? Answering these questions will also give us an idea of how robust our analysis is. In order to investigate this we have included in Fig 6 the RMS fit obtained after first removing the ripple in Fig 4 from the raw V$_{\rm CMB}$ data. The dip is still clearly present and it is therefore obvious that the presence or absence of the ripple in the raw data does not affect the fitting of the discrete components. This is most likely because the ripple components are small relative to the separation of the discrete components. It also indicates that the dip at H$_{\rm o}$ = 58 is produced entirely by the presence of the discrete components in the data. 

\begin{figure}[h,t]
\hspace{-0.5cm}
\vspace{-1.5cm}
\epsscale{0.9}
\includegraphics[width=8cm]{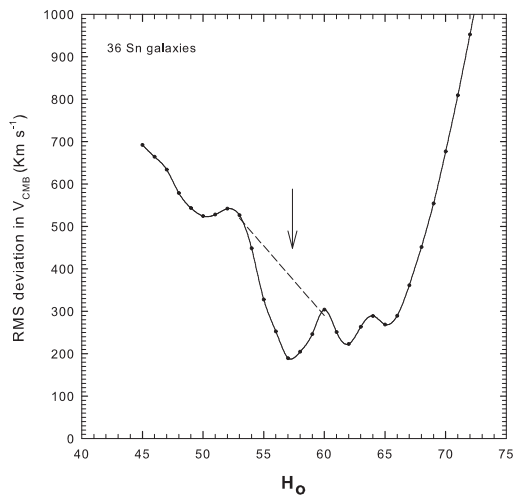}
%\includegraphics[width=9cm]{rmsplotrippleremovedfirst.eps}
%\plotone{fig5.eps}
%\plotone{vhslope40a.eps}
\caption{{Same as Fig 1 except the sinusoid in Fig 4 was removed from the V$_{\rm CMB}$ data before processing. \label{fig6}}}
\end{figure}
 
\begin{figure}[h,t]
\hspace{-0.8cm}
\vspace{-1.5cm}
\epsscale{0.9}
\includegraphics[width=9cm]{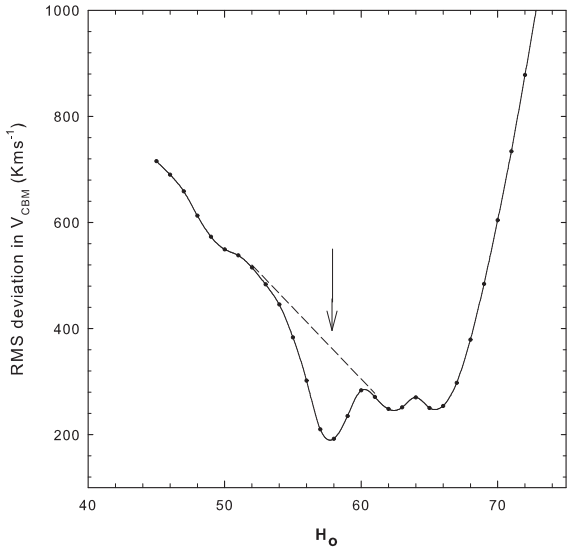}
%\includegraphics[width=9cm]{rmsmeannoshift.eps}
%\plotone{fig6.eps}
%\plotone{vhslope40a.eps}
\caption{{Mean of Fig 1 and Fig 6. \label{fig7}}}
\end{figure}

However, as can be seen in Fig 6 there are slight differences between this curve and the curve in Fig 1. It can be assumed that there are also small peculiar velocities and distance uncertainties present in the data. If these affect the analysis differently with and without the ripple being present they may introduce differences that are truly random. If so, their effect can be minimized by averaging the curves in Fig 1 and Fig 6. This will minimize the effect of random fluctuations but should not affect any real features. This has been done in Fig 7 where it is seen that while the dip at H$_{\rm o}$ = 58 is unaffected, some of the weaker features are reduced. We conclude that the dip at H$_{\rm o}$ = 58 is a real feature that is likely to have been produced by the presence of discrete components in the V$_{\rm CMB}$ data as we have argued previously. This conclusion is strengthened by the fact that a dip has been seen in other source samples at the same H$_{\rm o}$ = 58 Hubble slope.

\section{Conclusions}

We have identified what we believe are discrete components in the radial velocities of the 36 Type Ia Supernovae galaxies studied in the Hubble Key Project. These are defined by the same relation that also defines those identified previously by Tifft and by us in several other independent galaxy groups. As in our previous work this result is obtained for a Hubble constant of H$_{\rm o}$ = 58 kms$^{-1}$ Mpc$^{-1}$. We show that when these components are removed from the redshifts of the SNeIa galaxies there is evidence for a low-level sinusoidal oscillation superimposed on the Hubble flow. It is isotropic in nature and has a period of 40 Mpc.
%The oscillation has an amplitude that increases with distance at a rate of %$1.67\pm0.02$ km s$^{-1}$ Mpc$^{-1}$, such that $\Delta$H$_{\rm o}$/H$_{\rm o}$ %= $\pm0.029$ for H$_{\rm o}$ = 58.
We have been unable to identify any systematic effect that could have produced this observed oscillation in the Hubble flow. If one can be identified it will mean that the oscillation is not a real oscillation in the Hubble flow. However, it will also mean that the discrete components must then be real since they have to be removed before the selection effect is visible and this is unlikely to happen if the discrete components are random ones. If no selection effect can be identified that can explain the ripple, then both the discrete velocities and the ripple together become very difficult to explain by chance and these results could then have significant cosmological consequences.

\section{Acknowledgements}

I thank D.R. McDiarmid for continued support and helpful comments, and S.P. Comeau for assistance with the data analysis, and especially, for writing the computer program that made it possible to do RMS fitting of several parallel lines to the data simultaneously.

%\clearpage

%\end{document}
%\clearpage

\end{document}